\documentclass[aps,prl,twocolumn,floats,epsfig,showpacs]{revtex4-1}

\usepackage{bbm}
\usepackage{amssymb}
\usepackage{ifpdf}
\usepackage{color}
\usepackage{graphicx}
\usepackage{amsmath}
\usepackage{amsfonts}
\usepackage{hyperref}
\usepackage{dcolumn}
\usepackage{bm}
\usepackage{xcolor}
\usepackage{subfigure}
\usepackage{verbatim}

\newcommand{\Z}{{\mathbb{Z}}}

\definecolor{red}{rgb}{0.7,0,0}
\definecolor{green}{rgb}{0.,0.35,0.}
\definecolor{blue}{rgb}{0.2,0.2,0.7} 
\definecolor{black}{rgb}{0.15,0.15,.15}

\begin{document}

\title{Tensor Networks for Lattice Gauge Theories \\
and Atomic Quantum Simulation}

\author{E. Rico$^1$, T. Pichler$^1$, M. Dalmonte$^{2,3}$, P. Zoller$^{2,3}$, S. Montangero$^1$}
\affiliation{$^1$Institute for Quantum Information Processing, Ulm University, D-89069 Ulm, Germany}
\affiliation{$^2$Institute for Theoretical Physics, Innsbruck University, A-6020 Innsbruck, Austria}
\affiliation{$^3$Institute for Quantum Optics and Quantum Information of the Austrian Academy of Sciences, A-6020 Innsbruck, Austria}

\begin{abstract}

We show that gauge invariant quantum link models, Abelian and non-Abelian, can be exactly described in terms of tensor networks states. Quantum link models represent an ideal bridge between high-energy and cold atom physics, as they can be used in cold-atoms in optical lattices to study lattice gauge theories. In this framework, we characterize the phase diagram of a (1+1)-d  quantum link version of the Schwinger model in an external classical background electric field: the quantum phase transition from a charge and parity ordered phase with non-zero electric flux to a disordered one with a net zero electric flux configuration is described by the Ising universality class.
\end{abstract}

\maketitle

Lattice gauge theories (LGTs) play a key role in our understanding of quantum many body systems. In high energy physics they provide non-perturbative approach to Abelian and non-Abelian continuum gauge models like quantum electrodynamics (QED) or quantum chromodynamics (QCD) \cite{LGT1,LGT2}. In condensed matter context, they emerge as the low-energy description of some strongly correlated quantum systems \cite{cond}. While Monte Carlo techniques provide a well-established and highly successful framework to simulate LGT for equilibrium phenomena, the problem of real-time evolution and overcoming the Fermion sign problem remain key challenges in this field \cite{Wiese2}. These questions can be addressed by both novel classical and quantum simulation techniques. In a classical context, quantum information theory has provided new insights into the efficient description of quantum many body systems, for example in terms of tensor networks \cite{Cirac}. It is the purpose of the present work to develop a description of a Hamiltonian formulation of LGT in terms of tensor networks, with emphasis on a natural implementation of gauge constraints in the formalism. This is of interest not only from the perspective of quantum many-body physics, in particular for low dimensional (1D) systems, but also in the ongoing quest to develop a quantum simulator for Abelian and non-Abelian LGTs with cold atoms in optical lattices \cite{NatureInsightReview}. The techniques developed in the present paper provide the basis for an efficient and reliable calculation for phase diagrams and real time dynamics (quenches) of simple lattice gauge models, which will allow the verification of the first generation of atomic quantum simulator of LGTs. We will demonstrate this below for a (1+1)-d  {\em quantum link} version of the Schwinger model representing a model of QED.

The efficient simulation of quantum states by classical methods and in particular parametrized by a set of tensor-networks has been a major effort in recent years \cite{Perez,Cirac}. On one hand, tensor networks are an exact description of ground states of paradigmatic Hamiltonians, e.g., 2D toric code that is an Ising gauge theory \cite{Kitaev,Ardonne}. On the other hand, this framework is at the core of many successful sign-problem free numerical tools \cite{White,DMRG} which have been successfully applied to LGT related problems \cite{Martin,Byrnes,Haegeman,Luca,Banuls,Silvi}. A particular class of tensor networks, called matrix product state (MPS) is a common description for one dimensional systems \cite{MPS}. In this context, we show that MPS and tensor networks in higher dimensions are exact descriptions of the Gauss' law constraint of quantum link models with Abelian and non-Abelian local symmetries and we use them to describe different phases that can appear in these models.
\begin{figure}[t]
\centering
\includegraphics[width=8.cm]{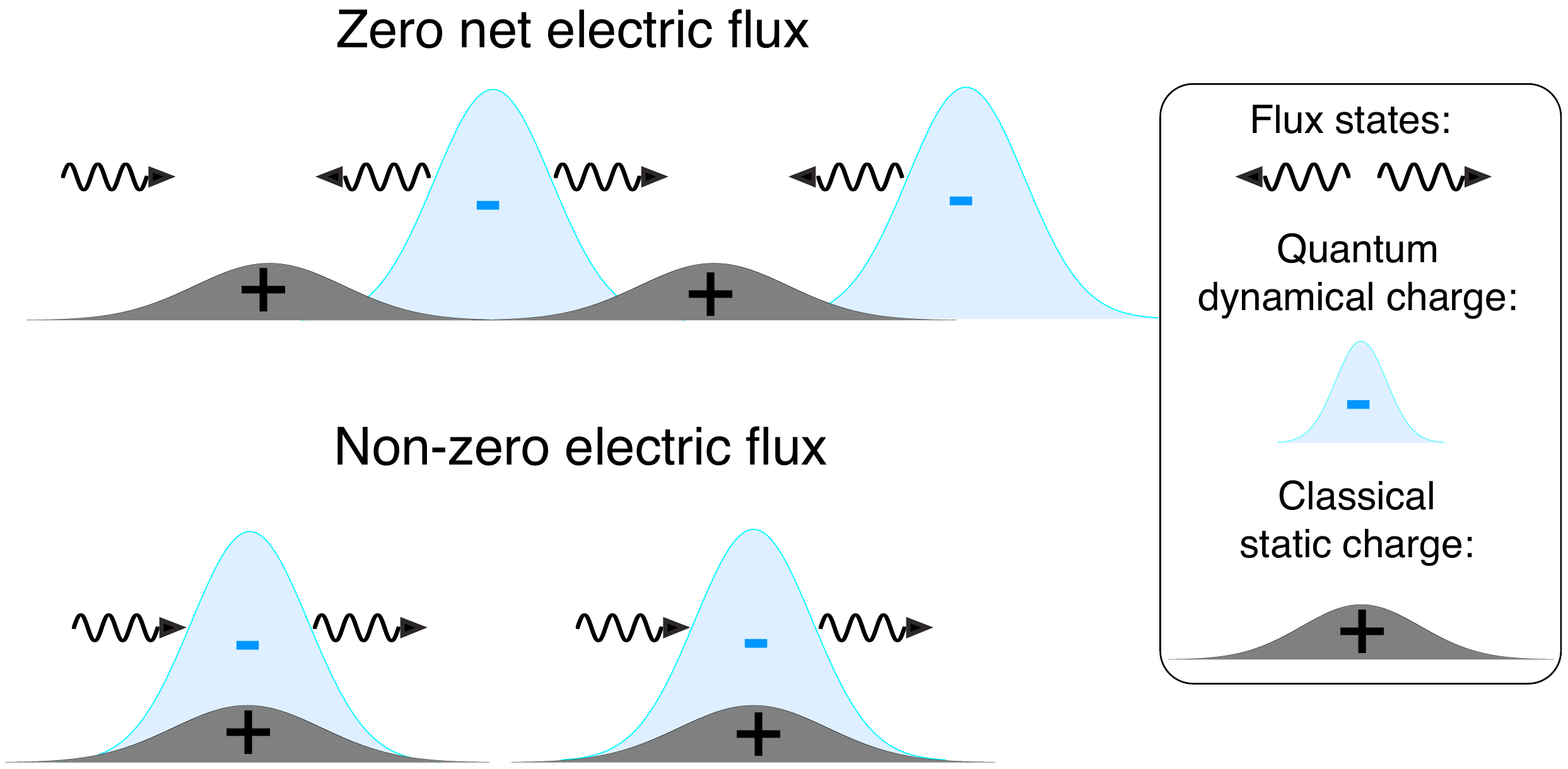}
\begin{caption}{\label{plot1} Ground state of the spin-$\frac{1}{2}$ quantum link model in the limiting cases of $|\mu| \gg |\epsilon|$ : in the upper (lower) panel the fermion and the gauge field states are represented for $\mu \ll \epsilon$ ($\mu \gg \epsilon$) resulting in zero electric flux, $\mathcal{E} = 0$, and C and P invariant state (non-zero electric flux, $\mathcal{E} \ne 0$, C and P symmetry broken).}
\end{caption}
\end{figure}

Quantum link models (QLM) provide an ideal playground to establish the connection between Abelian and non-Abelian LGT \cite{Horn,Orland,Wiese} and atomic lattice experiments \cite{Bloch2008}. In these models of LGT the dynamical gauge fields are represented by discrete degrees of freedom, e.g. spin operators, which have a natural mapping to a Hubbard-type Hamiltonian dynamics, which can be realized with cold atoms in optical lattices \cite{NatureInsightReview,Zohar,Banerjee,Tagliacozzo,Marcos,Hauke,Wiese2}.

The starting point for our discussion are LGTs in the Hamiltonian formulation, where gauge degrees of freedom $U_{x,x+1}$ are defined on links of a lattice, and are coupled to the matter ones $\psi_{x}$, defined on the vertices. In what follows, we specialize to a $U(1)$ quantum link model, although the theoretical analysis can be generalized to any gauge symmetry group $U(N)$ or $SU(N)$ and space-time dimension $d$ (see SI \cite{supp}). The simplest non-trivial Hamiltonian is of the form, $H=\sum_{x}  \psi_x^{\dag} U_{x,x+1} \psi_{x+1} +  \text{h.c.}$ which describes the coupling between the ``photon'' field $U_{x, x+1}$ and the electrons $\psi_x$. In the quantum link formulation, the gauge degrees of freedom are described by bilinear operators $U_{x,x+1}=c_{x,l} c^{\dagger}_{x+1,r}$ recasting the interaction term in a four-body Hubbard-type Hamiltonian. As we will see, this feature allows us to solve exactly, within the tensor network representation, the constraints imposed by the local symmetries of this model.

Quantum link models have two independent local symmetries:

\emph{(i)} Gauge models are invariant under local symmetry transformations. The local generators of these symmetries, $G_{x}$, commute with the Hamiltonian, $[H,G_{x}]=0$. Hence, $G_{x}$ are constant of motion or local conserved quantities, which constraint the \emph{physical} Hilbert space of the theory, $G_{x} |\text{phys}\rangle =0, \, \forall x$, and the total Hilbert space splits in a physical or gauge invariant subspace and a gauge variant or unphysical subspace: $\mathcal{H}_{\text{total}} = \mathcal{H}_{\text{phys}} \oplus \mathcal{H}_{\text{unphys}} $. In QED, this gauge condition is the usual Gauss' law. 

\emph{(ii)} The quantum link formulation of the gauge degrees of freedom introduces an additional constraint at every link, that is, the conservation of the number of link-particles, $c^{\dagger}_{x,l}c_{x,l}+c^{\dagger}_{x+1,r}c_{x+1,r} = N_{x,x+1}$. Hence, $[H,N_{x,x+1}]=0$ which introduce a second and independent local constraint in the Hilbert space.

In the following, first, we present the theoretical characterization of the local constraint \emph{(i)} and $\emph{(ii)}$ in terms of tensor networks. Secondly, we exploit this exact representation to implement a MPS-based approach which allows us to characterize the full phase diagram of non-trivial gauge invariant models. In particular, we study a quantum link version of the Schwinger model identifying the different phases and the universality class of the phase transition in the presence of a background field. 

\begin{figure*}[t]
\centering
\includegraphics[width=16.cm]{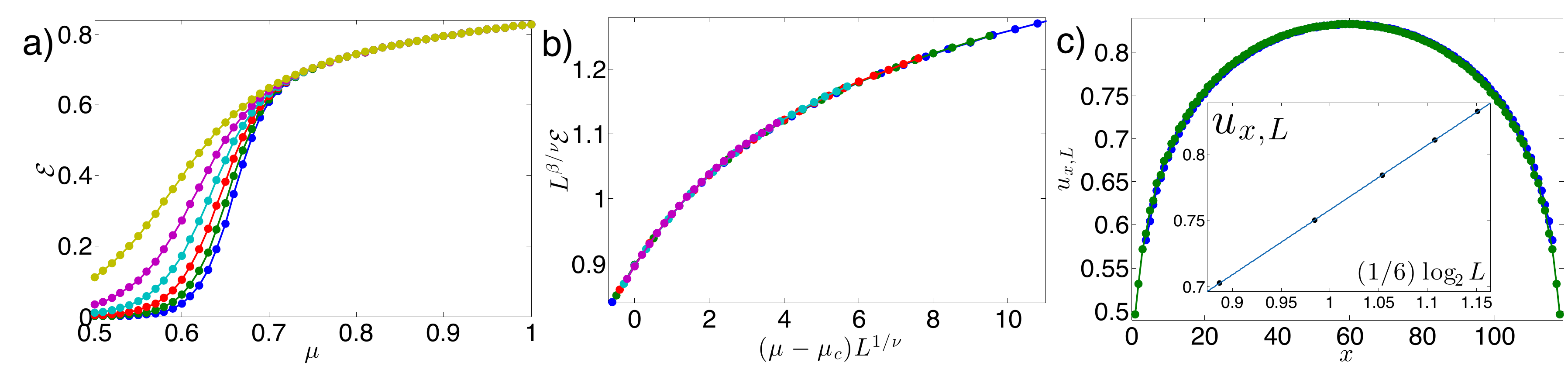}
\begin{caption}{\label{plot3} a) Electric flux $\mathcal{E}$ as a function of $\mu$ for $L=\{40,60,80,100,120,140\}$ from top to bottom, $S=1/2$ and $D=30$. b) Finite size scaling of the electric flux $\mathcal{E}$ shown in panel a, resulting in the critical point $\mu_{c} = 0.655\pm 0.003$ and critical exponents $\nu \sim 1$ and $\beta \sim \frac{1}{8}$. c) Uniform part of the entanglement entropy (green circles, first order approximation, i.e. $u_{x,L} = \frac{1}{2} \left( u_{x,L} + u_{x+1,L} \right)$, and blue squares, third order approximation~\cite{Entanglement}). Inset:  fit of $u_{x,L}$ as a function of the system size $\log L$: a linear fit results in the central charge $c = 0.49 \pm0.01$.}
\end{caption}
\end{figure*}

\paragraph{The gauge invariant model.-} Gauge theories in $(1+1)$ dimensions, and in particular the Schwinger model describing quantum electrodynamics in one space and one time dimension~\cite{Schwinger,Coleman1,Coleman2}, are non-trivial interacting models of fermions and gauge fields. They provide a playground to compute and understand many interesting phenomena with surprising analogies with non-abelian gauge theories in higher dimensions as, to name a few, the confinement of fermionic degrees of freedom and the appearance of a massive boson in the spectrum, chiral symmetry breaking through the axial anomaly, screening of external charges, and a topological $\theta$-vacuum. In particular, we consider  a $U(1)$ gauge invariant model in $(1+1)$ dimensions defined by the Hamiltonian
\begin{equation}
\begin{split}
H =  & \frac{g^{2}}{2} \sum_x  \left( E_{x,x+1}  - \left(-1 \right)^{x} E_{0} \right)^{2} \\
+&\mu \sum_x \left( -1 \right)^{x} \psi_x^{\dag}\psi_x -\epsilon \sum_{x}  \psi_x^{\dag} U_{x,x+1} \psi_{x+1} +  \text{h.c.},
\end{split}
\end{equation}
where $\psi_x$ are spin-less fermionic operators (matter fields with a staggered mass term $\mu$) living on the vertices of the one dimensional lattice, i.e., $\{\psi_x, \psi^{\dagger}_y \} = \delta_{x,y}$, usually denoted as {\it staggered} fermions~\cite{Kogut,Kogut1979}. The vacuum of the staggered fermions is given by a quantum state at half-filling describing the Fermi-Dirac sea. The bosonic operators $E_{x,x+1}$ and $U_{x,x+1}$ (electric and gauge field) live on the links of the one dimensional lattice, such that $\left[ E_{x,x+1} , U_{y,y+1}\right] = \delta_{x,y} U_{x,x+1}$. The coupling constant that measures the strength of the electric energy term is from now on set to one, i.e.,  $\frac{g^{2}}{2}=1$ while $\epsilon$ describes the interaction between the matter and gauge fields. Finally,  $E_{0}$ corresponds to a classical background field which at $E_{0}=\frac{1}{2}$, the ground state at every link is two-fold degenerate. In the Wilson formulation, the lattice Schwinger model has been numerically investigated using Monte Carlo techniques~\cite{Martin1,Martin2}, strong coupling expansion~\cite{Banks1976a,Banks1976b,Banks1976c} and MPS-based methods~\cite{Byrnes,Banuls}.

The quantum link \cite{Horn,Orland,Wiese} representation of the gauge degrees of freedom is given by the $SU(2)$ spin operators if we identify: $E_{x,x+1} \equiv S^{(z)}_{x,x+1}$ and $U_{x,x+1} \equiv S^{+}_{x,x+1}$. We use Schwinger bosons ($c_{x,l}$, $c_{x+1,r}$) to represent the spin algebra such that $U_{x,x+1} \equiv S^{+}_{x,x+1} = c_{x,l} c^{\dagger}_{x+1,r}$ where we have introduced a local set of states given by the occupation numbers of bosons on the right $(x,r)$, on the fermion $(x)$ and on the bosons on the left $(x,l)$ as follows $| n_{x,r} , n_{x} , n_{x,l} \rangle$. The number of bosons per link $N_{x,x+1}$ determines the representation of the spin. In this work, we use the two smallest integer and half-integer representations, i.e., $S=\frac{1}{2}$ for $N_{x,x+1}=1$ and $S=1$ for $N_{x,x+1}=2$.

With these definitions, the Hamiltonian is invariant under local $U(1)$ symmetry transformations, and also it is invariant under the discrete parity transformation $P$ and charge conjugation $C$ (see SI \cite{supp}). Due to the $\mathbb{Z}_{2}$ discrete nature of these symmetries, they can be broken in one dimensional systems, allowing critical points between a $CP$ broken phase and an unbroken one. The order parameter, the total electric flux, $\mathcal{E}=\sum_{x} \langle E_{x,x+1}\rangle /L=\sum_{x} \langle S^{(z)}_{x,x+1} \rangle /L$ locates the transition. It is zero in the disordered phase, non-zero in the ordered phase and changes the sign under the $C$ or $P$ symmetry, i.e., $^{P}\mathcal{E} = ^{C}\mathcal{E} =-\mathcal{E}$.

Representative states of the different phases appear at the strong coupling limit $|\mu| \gg |\epsilon|$ where the Hamiltonian is given by $H_{\text{str}} = \mu \sum_{x} \left( -1 \right)^{x} \psi^{\dagger}_{x} \psi_{x}$ (sketched in Fig.~\ref{plot1}). For $\mu \gg \epsilon$, due to the gauge invariance, the Hamiltonian has two possible ground states where the configuration of the fermions is staggered (leftmost occupied site) and the configuration of the bosons is also staggered with two possible patterns. This phase is two-fold degenerate, the vacuum states break charge and parity symmetry and they have non-zero electric flux. For $\mu \ll \epsilon$, the vacuum is unique and the staggered fermion has the rightmost site occupied. This phase is $C$ and $P$ symmetric and it has a net zero electric flux.

\paragraph{The ``physical'' subspace-} The number of bosons per link $N_{x,x+1}=N$ is a local conserved quantity of the model written in terms of Schwinger bosons.  Due to gauge invariance of the model, i.e., $[H,G_{x}]=0$, the gauge generator of the local $U(1)$ symmetry $G_x$ is a second local conserved quantity. The usual convention is to define the ``physical'' subspace as the one that fulfills $G_{x}|\text{physical}\rangle=0, \, \forall x$ \cite{Kogut}. In a quantum link model, we can solve the gauge invariance or Gauss' law locally, that is, in terms of the Schwinger bosons the constraint is given by
\begin{equation}
c^{\dagger}_{x,r} c_{x,r} + \psi^{\dagger}_{x} \psi_{x} + c^{\dagger}_{x,l} c_{x,l}  \big|_{\text{phys}} = N -\frac{\left( -1\right)^{x} - 1}{2}.
\end{equation}
Due to this feature, we can show that the gauge invariant condition and the conserved number of bosons per link can be written exactly in a MPS form. Indeed, the Gauss projection can be done locally defining the local Hilbert space $\{|s_{x} \rangle\}$, while the link representation is implemented by the product between the MPS matrices. Recently, the action of global symmetries on MPS-like wave function has been considered\cite{Singh,Bauer,Weichselbaum}, what follows can be seen as the counterpart of this for local (gauge) symmetries.

For concreteness, we build the MPS for a case with $S=\frac{1}{2}$ on the link, but a similar discussion can be carried out for any representation $S$, gauge symmetry group, Abelian or non-Abelian, and space-time dimensions for the QLM (see SI \cite{supp}).

For $N=1$ bosons per link, there are just three local gauge invariant states $| n_{x,r} , n_{x} , n_{x,l} \rangle$ which configurations depend in the site, if it is odd $(n_{2x-1,r} + n_{2x-1} + n_{2x-1,l} = 2)$ or even $(n_{2x,r} + n_{2x} + n_{2x,l} = 1)$. Being a spin-$1/2$ the representation of the quantum link variable implies that on every link, there is only one boson present, i.e., $n_{x,l} + n_{x+1,r}=1$. These two conditions are fulfilled if the wave function has a general MPS form
\begin{equation}
\begin{split}
|\text{phys} \rangle = &\sum_{s_{1}, \cdots , s_{x}, \cdots } a\left( s_{1}, \cdots, s_{x}, \cdots \right) \\
&\text{Tr} \left\{ A[s_{1}] \cdots A[s_{x}] \cdots  \right\} |s_{1}, \cdots, s_{x}, \cdots \rangle
\end{split}
\end{equation}
with $A[1]=\begin{pmatrix}0 & 0 \\ 1 & 0  \end{pmatrix}$, $A[2]= \begin{pmatrix}1 & 0 \\ 0 & 0\end{pmatrix}$, $A[3]= \begin{pmatrix}0 & 1 \\ 0 & 0\end{pmatrix}$, this MPS structure codifies both the gauge invariance and the representation of the link variable; $a\left( s_{1}, \cdots, s_{x}, \cdots \right)$ is a general amplitude, in principle non-local that could also be represented by a MPS.

\paragraph{MPS as a variational set.-} To get the thermodynamical properties of this model, we use an imaginary time evolution algorithm with a MPS decomposition of the ground state \cite{Vidal, Daley}. We show results for chains with up to $L=140$ sites and bond dimension $D$ up to $30$.

We use open boundary conditions (see Fig. \ref{plot1}) fixing the occupation of the first boson to one, $\langle c^{\dagger}_{1,r} c_{1,r} \rangle = 1$, and the occupation of the last boson to zero, $\langle c^{\dagger}_{L,l} c_{L,l} \rangle = 0$. With these boundary conditions, we  observe the transition between both phases and we avoid the double degeneracy of the $CP$ broken phase.

The parameter that controls the transitions between the different phases is the staggered mass $\mu$ of the fermions. From the behavior of the order parameter $\mathcal{E}$ we extract an estimate of the critical point and of the critical exponents. Due to the $\mathbb{Z}_{2}$ parity and charge conjugation symmetries, the critical point belongs to the Ising universality class, as confirmed by the following numerical analysis. Indeed, the finite size scaling hypothesis predicts the order parameter $ \mathcal{E} $ behavior close to a critical point $\mu_{c}$ as: $ \mathcal{E}  \sim L^{-\beta/\nu} f\left[ L^{1/\nu} \left( \mu-\mu_{c} \right) \right]$, with a scaling function $f(x)$ and critical exponents $\beta$ and $\nu$~\cite{Fisher,Silvi}. In Fig. \ref{plot3}, we show how the behavior of the electric field density goes to zero in the disordered phase, when $\mu \ll \epsilon$, and to a finite value in the ordered phase, when $\mu \gg \epsilon$. We computed the critical value of the staggered mass $\mu_{c} = 0.655\pm 0.003$ and found critical exponents compatible with $\nu \sim 1$ and $\beta \sim \frac{1}{8}$. 

We use the entanglement entropy as an order parameter to detect the phase transition~\cite{Amico}. The first thing to notice is the oscillatory behavior of the entanglement entropy due to the constraint on the number of bosons per link (see SI \cite{supp}). To decouple the uniform from the oscillatory behavior, we define two auxiliary functions, $S_{x,L}=u_{x,L}+\left( -1\right)^{x} o_{x,L}$. Conformal field theory analysis proved that the uniform part of the entanglement entropy is given by $u_{x,L} =  \frac{c}{6} \log{ \left[ \frac{2L}{\pi} \sin \left( \frac{\pi x}{L} \right) \right] } + a $ where $c$ and $a$ are constants to be fitted~\cite{Calabrese,Entanglement}. In the continuum limit $c$ corresponds to the central charge and determines the universality class of the model. In our calculations, we obtain $c = 0.49 \pm0.01$, consistent with an Ising transition (see Fig. \ref{plot3} and SI \cite{supp}).  In conclusion, the central charge is in very good agreement with an Ising universality class at the phase transition.

\begin{figure}[t]
\centering
\includegraphics[width=8.cm]{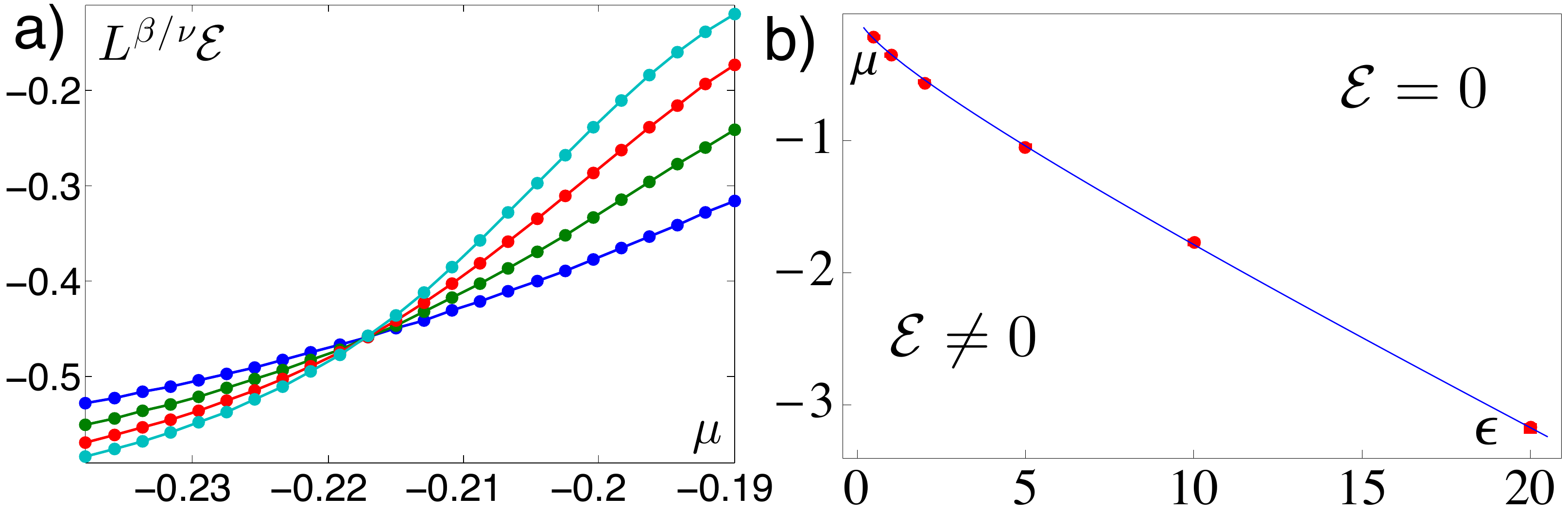}
\begin{caption}{\label{plot4} 
a) Electric flux $\mathcal{E}$ for $S=1$, $L=\{40,60,80,100\}$, $D=30$ and $\epsilon=0.5$ as a function of $\mu$. The estimated critical point is $\mu_{c} = -0.2173\pm 0.0005$. b) Phase diagram
of the $S=1$ representation. The critical line is fitted via  $\mu_{c} \sim \mu_{0} + \mu_{\frac{1}{2}} \sqrt{\epsilon}  + \mu_{1} \epsilon$ resulting in $\mu_{0}= - 0.04 \pm 0.03$, $\mu_{\frac{1}{2}}= - 0.20 \pm 0.03 $ and $\mu_{1}=-0.113 \pm 0.005 $.}
\end{caption}
\end{figure}

Once we have characterized the behavior of the quantum link model with $S=1/2$ representation on the links, we compare it with the $S=1$ one case (see Fig. \ref{plot4} and SI \cite{supp}). The main difference in the Hamiltonian of both models is that with the integer representation, we apply a background electric field $E_{0}=\frac{1}{2}$. With this value of the background field, the Hamiltonian is still C and P invariant, nonetheless the vacuum can spontaneously break these symmetries as in the $S=1/2$ link representation. We find critical line $(\mu,\epsilon)$ fitted as $\mu_c  \sim \mu_{0} + \mu_{\frac{1}{2}} \sqrt{\epsilon}  + \mu_{1} \epsilon$, with $\mu_{0}= - 0.04 \pm 0.03$, $\mu_{\frac{1}{2}}= - 0.20 \pm 0.03 $ and $\mu_{1}=-0.113 \pm 0.005 $, belonging again to the Ising universality class. Hence, the thermodynamical properties and phase diagram of a quantum link model with a half integer link representation are the same as a quantum link model with integer representation in a classical background field $E_{0}=\frac{1}{2}$.

\paragraph{Observability in synthetic systems.-} The (1+1)-d quantum link model investigated in this manuscript has been discussed in relation to different atom, ion and solid state platforms~\cite{Zohar, Banerjee, Tagliacozzo, Hauke, Marcos}. The figure of merit to access the ground state physics of the model is the order parameter $\mathcal{E}$ that can be measured in the different platforms as described below. 

Different implementation schemes using ultra cold atomic gases have introduced various ways of realizing the Abelian gauge fields. In Ref.~\cite{Banerjee}, the spin degrees of freedom are realized in terms of Schwinger bosons in arrays of double well potentials, $S^{z}_{x,x+1} = \frac{1}{2}(n^{(a/b)}_{x+1}-n^{(a/b)}_{x})$. Here, two species $a$ and $b$ are used for odd/even - even/odd links, with respective number operators $n^{a/b}_{x}$, and the spin representation is given by the number of bosons in each double well (i.e., one and two bosons per well for $S=1/2, 1$ representations, respectively). In this case, the expectation value of the order parameter can be measured in two possible ways. The first one is, to employ the recently developed  quantum gas microscope~\cite{Bakr, Weitenberg} to perform {\it in situ} imaging measurement of the bosonic particle distribution. Since in real experiments local parity $(-1)^{n^\alpha}$ is accessible, this provides an exact measure of the local value of $S^{(z)}_{x, x+1}$ as long as $S\leq 3/2$ is considered. Alternatively, one can utilize band mapping techniques to measure the difference between the number of bosons on each side of the double well: this provides a global probe accessing directly the order parameter $\mathcal{E}$. Within this setup, for $S>1/2$, the background electric field can be implemented by applying a local off-set $\lambda \sum_{x\;\textrm{even}} n^{(a,b)}_x$ to the double wells, created by, e.g., a small imbalance of the superlattice potential. The corresponding $E_{0}$ value reads $E_{0} = \lambda /g^2$. Notice that, as $g^2$ can be tuned to small values, $E_{0}$ can reach large values within the perturbative regime assumed in Ref.~\cite{Banerjee}.

Other implementation schemes realize spin degrees of freedom by means of different internal states of an atom sitting on the link sites~\cite{zohar2013, stannigel2013}. In these cases, the global order parameter can be accessed via a Stern-Gerlach-type measurement, where one can selectively measure the global occupation number of each of the spin states.

\paragraph{Conclusions.-} We have shown that MPS and tensor networks in higher dimensions are exact descriptions of the Gauss' law and the ``physical'' gauge invariant subspace of quantum link models with Abelian and non-Abelian local symmetries. Here we have characterized the thermodynamical properties and phase diagram of a one-dimensional $U(1)$ quantum link model, concluding that the model with half-integer link representation has the same physical properties as the model with integer link representation in a classical background electric field $E_{0}=\frac{1}{2}$. Moreover, we have shown how the phase diagram can be probed in various synthetic systems, and how the electric background field can be engineered in cold atomic gases. This work constitute a relevant step towards a systematic understanding of lattice gauge theories in low-dimensional systems by employing numerical variational methods, and which could serve as a ideal benchmark for atomic platforms for either Abelian and non-Abelian lattice gauge theories.

We thank D. Banerjee, T. Calarco, P. Silvi, U.-J. Wiese for stimulating discussions.
We acknowledge support from EU through SIQS, the ERC Synergy grant UQUAM, from the German Research Foundation via the SFB/TRR21 and the BW-grid for computational resources, and from the Austrian Science Fund FWF (SFB FOQUS F4015-N16).

\cleardoublepage

\section{Appendix}

\subsection{Tensor networks and gauge invariance in quantum link models.-} 

\begin{figure}[th]
\centering
\includegraphics[width=8cm]{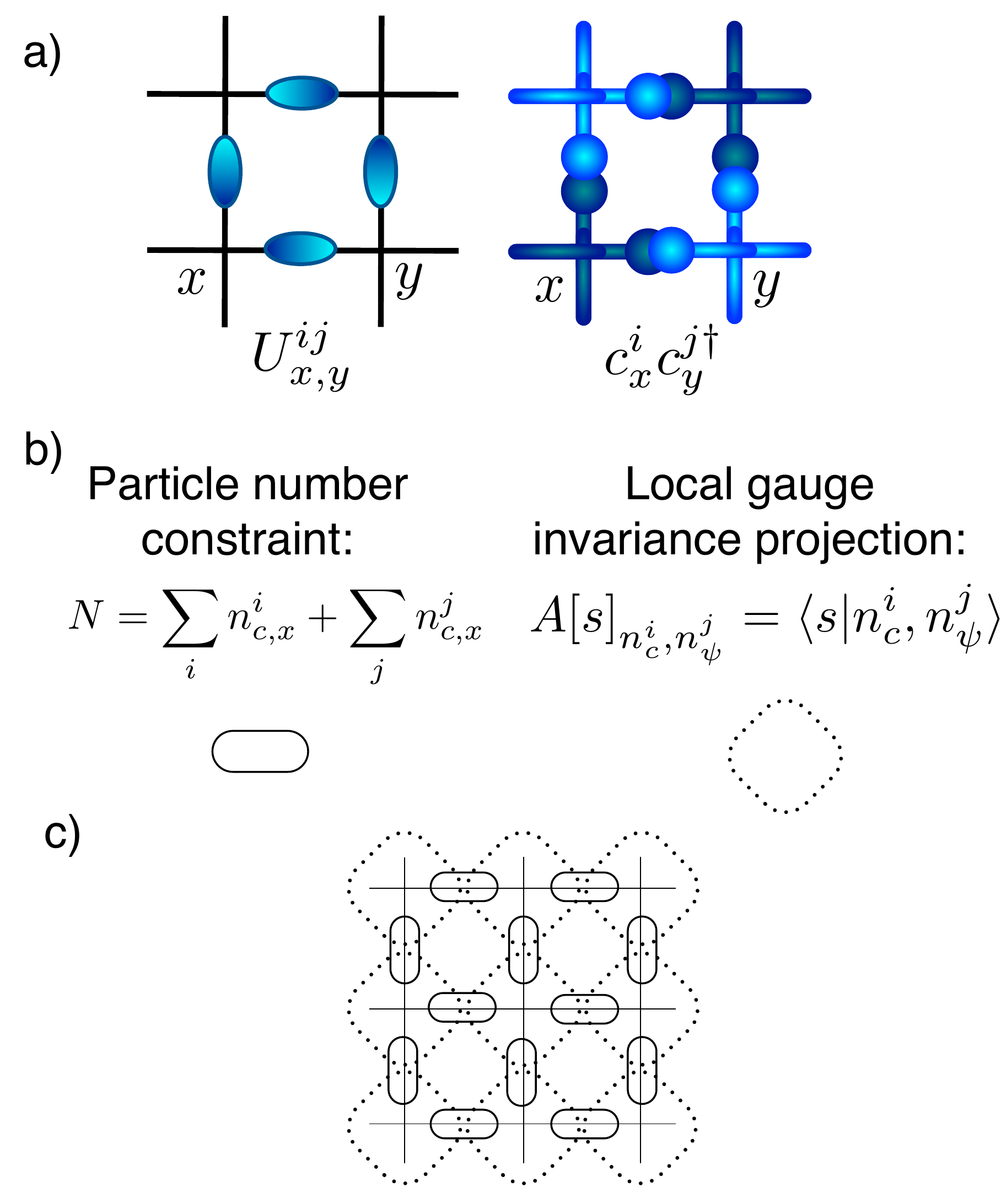}
\begin{caption}{\label{app1} a) General structure of a fermionic representation of a quantum link model with the link operators $U^{ij}_{x,y}=c^{i}_{x}c^{j\dagger}_{y}$. b) The two constraints, particle number on the links and gauge invariance, that have to be enforced in the ``physical'' Hilbert space of a quantum link model. c) Generalized tensor network formulation of the ``physical'' Hilbert space of a quantum link model.}
\end{caption}
\end{figure}

In the following, we show that the MPS decomposition that we have just presented in the main text for a $U(1)$ quantum link model in $(1+1)$ dimensions can be extended to $U(N)$ and $SU(N)$ quantum link models in higher dimensions. The case of orthogonal groups $O(N)$ is also subjected to a similar description. We follow the quantum link description from \cite{Brower}, starting with the degrees of freedom (gauge and matter fields) and showing how gauge invariance and link representation can be formulated with a tensor network generalization of the MPS.

In a fermionic Schwinger representation of a non-Abelian $U(N)$ quantum link model, the gauge operators $U^{ij}_{x,y}$ that live on the links $\langle x, y \rangle$ of a $d$ dimensional lattice, with color indices $i,j$ is expressed as a bilinear of fermionic operators, $U^{ij}_{x,y}=c^{i}_{x}c^{j\dagger}_{y}$. In this link representation, the number of fermions per link is a constant of motion $N_{x,y} = \sum_{i} c^{i\dagger}_{y} c^{i}_{y} + c^{i\dagger}_{x} c^{i}_{x}=N$. In models with matter, at every vertex $x$ of the lattice, there is a set of fermionic modes $\psi^{i}_{x}$ with color index $i$.

Within this representation, we can define left and right generators of the $SU(N)$ symmetry, $L^{a}_{x,y}=\sum_{i,j} c^{i\dagger}_{x} \lambda^{a}_{i,j} c^{i}_{x}$ and $R^{a}_{x,y}=\sum_{i,j} c^{i\dagger}_{y} \lambda^{a}_{i,j} c^{i}_{y}$, with $\lambda^{a}_{ij}$ the group structure constants. Hence, the non-Abelian generators of the gauge symmetry is given by $G^{a}_{x}=\sum_{i,j} \psi^{i\dagger}_{x} \lambda^{a}_{ij} \psi^{j}_{x} + \sum_{k} \left[ L^{a}_{x,x+\hat{k}} +R^{a}_{x-\hat{k},x} \right]$, with $\hat{k}$ the different directions in the lattice. There are also similar expressions for the Abelian part of the group $G_{x}$ \cite{Brower}.

Once we have defined the gauge generators, we define the ``physical'' Hilbert subspace as the one that is annihilated by every generator, i.e., $G_{x}|\text{phys}\rangle = G^{a}_{x}|\text{phys}\rangle = 0$, for all $x$ and $a$. A particular feature of quantum link models is that, being these operator of bosonic nature (they are bilinear combinations of fermionic operators), the spatial overlap between operators at different vertices $x$ or $y$ is zero, even between nearest neighbors, i.e., $G^{a}_{x} \cap G^{b}_{y} =0$, $\forall a,b$ and $x \neq y$. In this way, we can define locally a projection $A[s_{x}]$ on the ``physical'' subspace $\{|s_{x}\rangle \}$ with $A[s]_{n_{c},n_{\psi}} = \langle s | n^{i}_{c},n^{j}_{\psi} \rangle$, where $n^{i}_{c},n^{j}_{\psi}$ is some configuration of occupation of fermionic modes $c^{i}$ and $\psi^{j}$.

Finally, the fermionic number on the link is ensured by the product of the nearest neighbor projectors $A[s_{x}]$, being non-zero only when $N=\sum_{i} n^{i}_{c,y} + n^{i}_{c,x}$.

As an example, it is straightforward to check the non-Abelian case $U(2)$ in one spatial dimension where the gauge invariant subspace is spanned by the basis:
\begin{equation}
\begin{split}
&|1\rangle = |0, \uparrow \downarrow, 0 \rangle \\
& |2\rangle = \frac{1}{\sqrt{2}} \left[ | \uparrow, \downarrow, 0 \rangle - | \downarrow, \uparrow, 0 \rangle \right] \\
&|3\rangle= \frac{1}{\sqrt{2}} \left[ | \uparrow, 0, \downarrow \rangle - | \downarrow, 0,  \uparrow \rangle \right]  \\
& |4\rangle = \frac{1}{\sqrt{2}} \left[ | 0 , \uparrow, \downarrow \rangle - | 0 , \downarrow, \uparrow \rangle \right] 
\end{split}
\end{equation}
and $A[1]=\begin{pmatrix}0 & 0 \\ 1 & 0  \end{pmatrix}$, $A[2]= \begin{pmatrix}1 & 0 \\ 0 & 0\end{pmatrix}$, $A[3]= \begin{pmatrix}0 & 1 \\ 0 & 0\end{pmatrix}$ and $A[4]= \begin{pmatrix}0 & 0 \\ 0 & 1\end{pmatrix}$

\subsection{Symmetries of the U(1) quantum link model}

In this section, we briefly review the basic symmetry properties of the $U(1)$ quantum link model of eq. (1) of the main text. 
\begin{enumerate}
\item As in any gauge theory, the Hamiltonian is invariant against local symmetry transformations. In this case, it commutes with the infinitesimal $U(1)$ gauge generators
\begin{eqnarray}
&&G_x = \psi_x^\dagger \psi_x + \frac{1}{2} \left[(-1)^x - 1\right] 
- E_{x,x+1} + E_{x-1,x}.
\end{eqnarray}

\item The parity transformation P is implemented as
\begin{eqnarray}
&&^{P}\psi_x = \psi_{-x},
~~~^{P}\psi_x^{\dag} = \psi_{-x}^{\dag}, \nonumber \\
&&^{P}U_{x,x+1} = U_{-x-1,-x}^{\dag},
~~^{P}E_{x,x+1} = - E_{-x-1,-x},
\end{eqnarray}

\item while charge conjugation C acts as
\begin{eqnarray}
&&^{C}\psi_x = (-1)^{x+1}\psi_{x+1}^{\dag},
~~~^{C}\psi_x^{\dag} = (-1)^{x+1} \psi_{x+1}, \nonumber \\
&&^{C}U_{x,x+1} = U_{x+1,x+2}^{\dag},
~~^{C}E_{x,x+1} = - E_{x+1,x+2}.
\end{eqnarray}

\item For $m = 0$ the Hamiltonian also has a $\Z(2)$ chiral symmetry which shifts all fields by one lattice spacing,
\begin{eqnarray}
&&^{\chi}\psi_x = \psi_{x+1},
~~~^{\chi}\psi_x^{\dag} = \psi_{x+1}^{\dag}, \nonumber \\
&&^{\chi}U_{x,x+1} = U_{x+1,x+2},
~~^{\chi}E_{x,x+1} =  E_{x+1,x+2}.
\end{eqnarray}
However, this symmetry is explicitly broken when one imposes the Gauss law $G_x|\Psi\rangle = 0$. 

\end{enumerate}

\begin{figure}[t]
\centering
\includegraphics[width=8.cm]{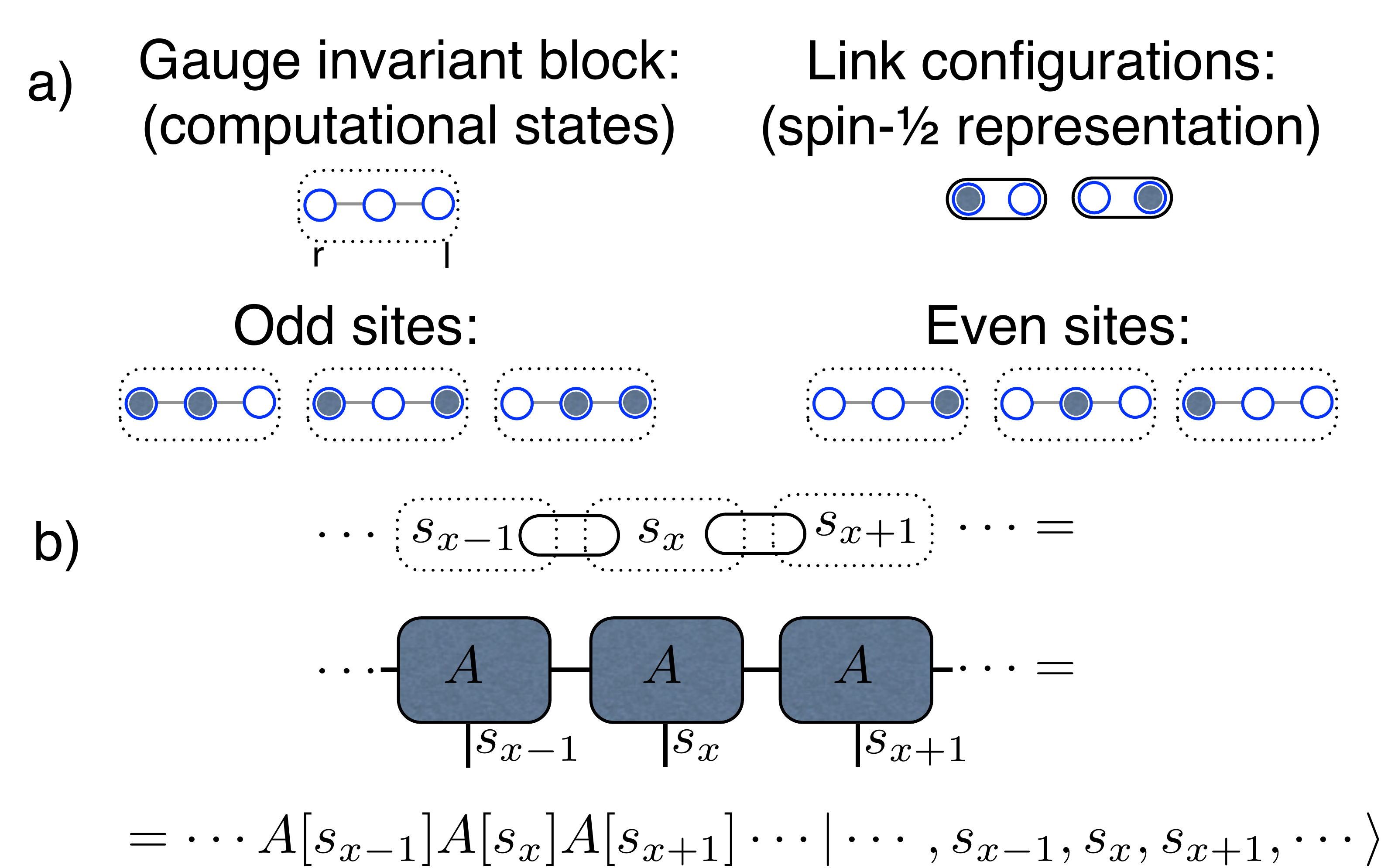}
\begin{caption}{\label{plot2} a) Gauge invariant and link configurations for $U(1)$ quantum link model with spin-$1/2$ representation (notation as in Fig.~1 main text). b) Exact matrix product state description of the gauge constraint and representation on the links.}
\end{caption}
\end{figure}

\subsection{Physical or gauge invariant states.-} 

\begin{figure}[th]
\centering
\includegraphics[width=8.cm]{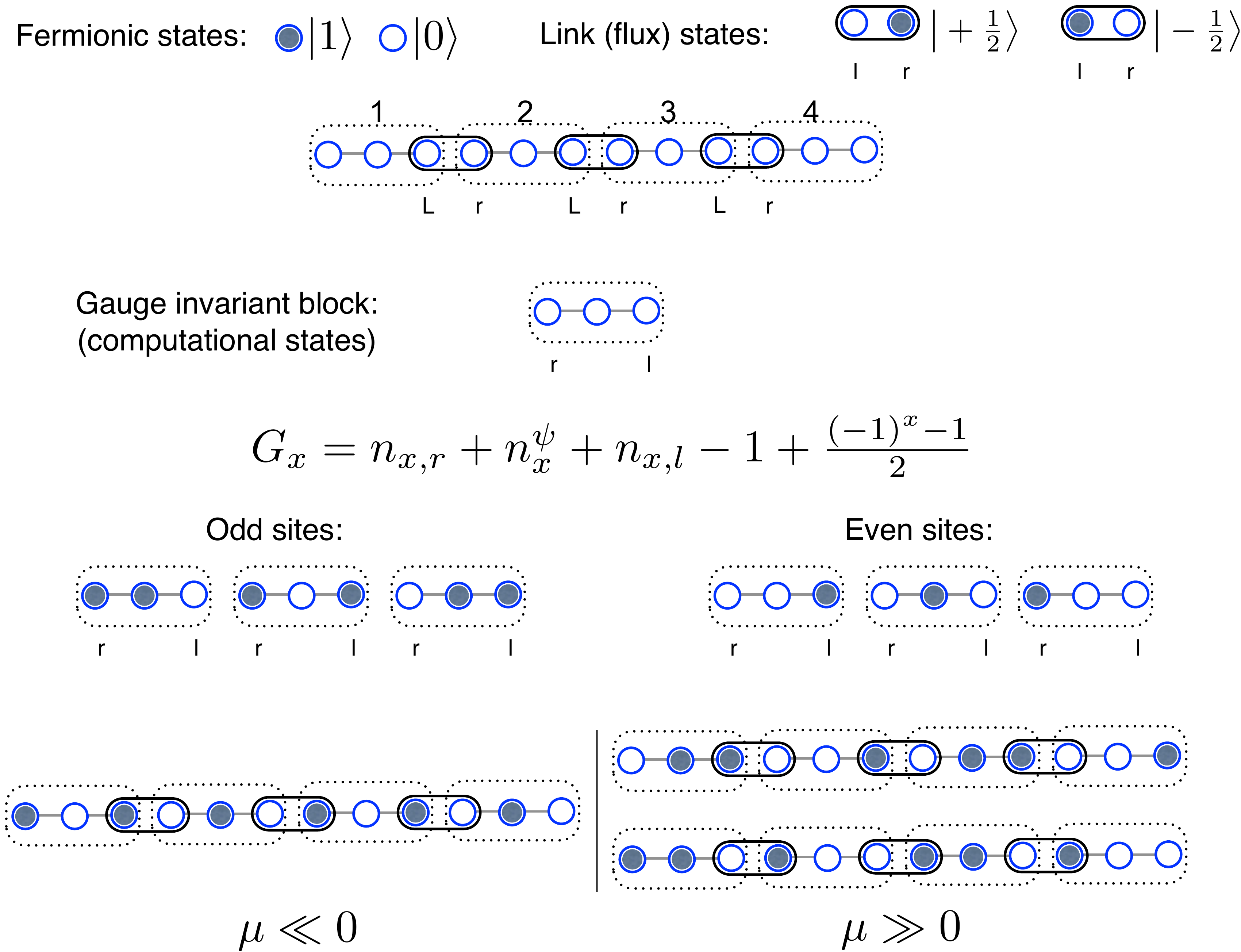}
\begin{caption}{\label{app2} Fermion and boson configurations. C and P invariant candidate vacuum state of the spin-$\frac{1}{2}$ model for $\mu \ll 0$. Competing vacua of the spin-$\frac{1}{2}$ model, which are C and P partners of each other for $\mu \gg 0$.}
\end{caption}
\end{figure}

\begin{figure*}[th]
\includegraphics[width=\textwidth]{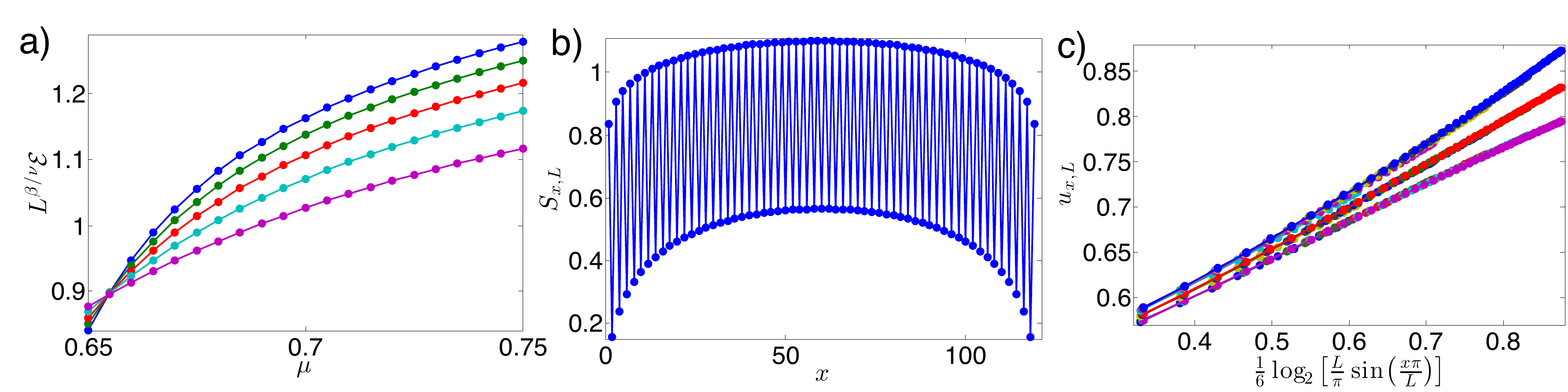}
\caption{Results for the model with spin-$1/2$ on the links: a) Electric flux density $\mathcal{E}$ as a function of the parameter $\mu$ for systems length $L=\{40, 60,80, 100, 120 \}$. b) Entanglement entropy at the phase transition. c) Uniform part of the entanglement entropy as a function of the scaling function $\frac{1}{6} \log_{2}\left[ \frac{L}{\pi} \sin \left( \frac{x \pi}{L} \right) \right]$ for several lengths $L \in \{ 40,60,80,100,120\}$ and parameters $\mu \in \{ 0.66, 0.655, 0.65\}$. The maximum overlap happens at the critical point $\mu_{c} \sim 0.655$.}
\label{app3}
\end{figure*}

The local states are given by occupation numbers of bosons on the right $(x,r)$, on the fermion $(x)$ and on the bosons on the left $(x,l)$ as follows $| n_{x,r} , n_{x} , n_{x,l} \rangle$. 

\subsubsection{Spin-1/2 on the link: $N=1$}

For $N=1$, the gauge invariant states are different, depending on the sites, if they are odd, $(2x-1)$, or even, $(2x)$, and they are given by
\begin{equation}
\begin{split}
(2x-1):  ~& | 1_{o} \rangle = |0,1,1 \rangle; ~ | 2_{o} \rangle = |1,0,1 \rangle; \\
& | 3_{o} \rangle = |1,1,0 \rangle;  \\
(2x):  ~ &| 1_{e} \rangle = |1,0,0 \rangle; ~ | 2_{e} \rangle = |0,1,0 \rangle; \\
& | 3_{e} \rangle = |0,0,1 \rangle; 
\end{split}
\end{equation}

With these definitions, the strong coupling limit Hamiltonian, i.e., $|\mu| \gg |\epsilon|$, is given by
\begin{equation}
\begin{split}
&H_{\text{str}} = \mu \sum_{x} \left( -1 \right)^{x} \psi^{\dagger}_{x} \psi_{x}  \\
&+ 2U \sum_{x} \left( c^{\dagger}_{x,l} c_{x,l} - \frac{1}{2} \right) \left( c^{\dagger}_{x+1,r} c_{x+1,r} - \frac{1}{2} \right) 
\end{split}
\end{equation}
where $U \gg 1$ constraints the state on the links to have just one boson.

For $\mu \gg 0$ and due to the gauge invariance, this Hamiltonian has only two possible ground states where the configuration of the fermions is staggered 1-0-1-0 and the configuration of the bosons is also staggered by with two possible patterns: 0-1-0-1-0-1-0-1 or 1-0-1-0-1-0-1-0. This phase is two fold degenerated and the vacuum states break charge and parity symmetry.

For $\mu \ll 0$, the staggered fermion configuration is given by 0-1-0-1 and the bosons configuration is unique, given by 1-1-0-0-1-1-0-0. This phase has a unique vacuum state that is C and P symmetric.

In the Fig. \ref{app3}, we can see the behavior of the electric flux density $\mathcal{E}$ as a function of the parameter $\mu$, with an estimate of the critical point $\mu_{c} = 0.655\pm 0.003$. Also, we can appreciate the oscillatory behavior of the entanglement entropy, due to the constraint on the number of bosons per link. To decouple the uniform from the oscillatory behavior of the entanglement entropy, we define two auxiliary functions such that $S_{x,L} = u_{x,L} + (-1)^{x} o_{x,L}$, where $u$ describes the uniform part and $o$, the oscillatory one. The uniform part follows the scaling relation $u_{x,L}=\frac{c}{6} \log_{2}\left[ \frac{L}{\pi} \sin \left( \frac{x \pi}{L} \right) \right]$, with an estimation of $c=0.49 \pm 0.01$ which corresponds to an Ising universality class.

\begin{figure}[th]
\centering
\includegraphics[width=8cm]{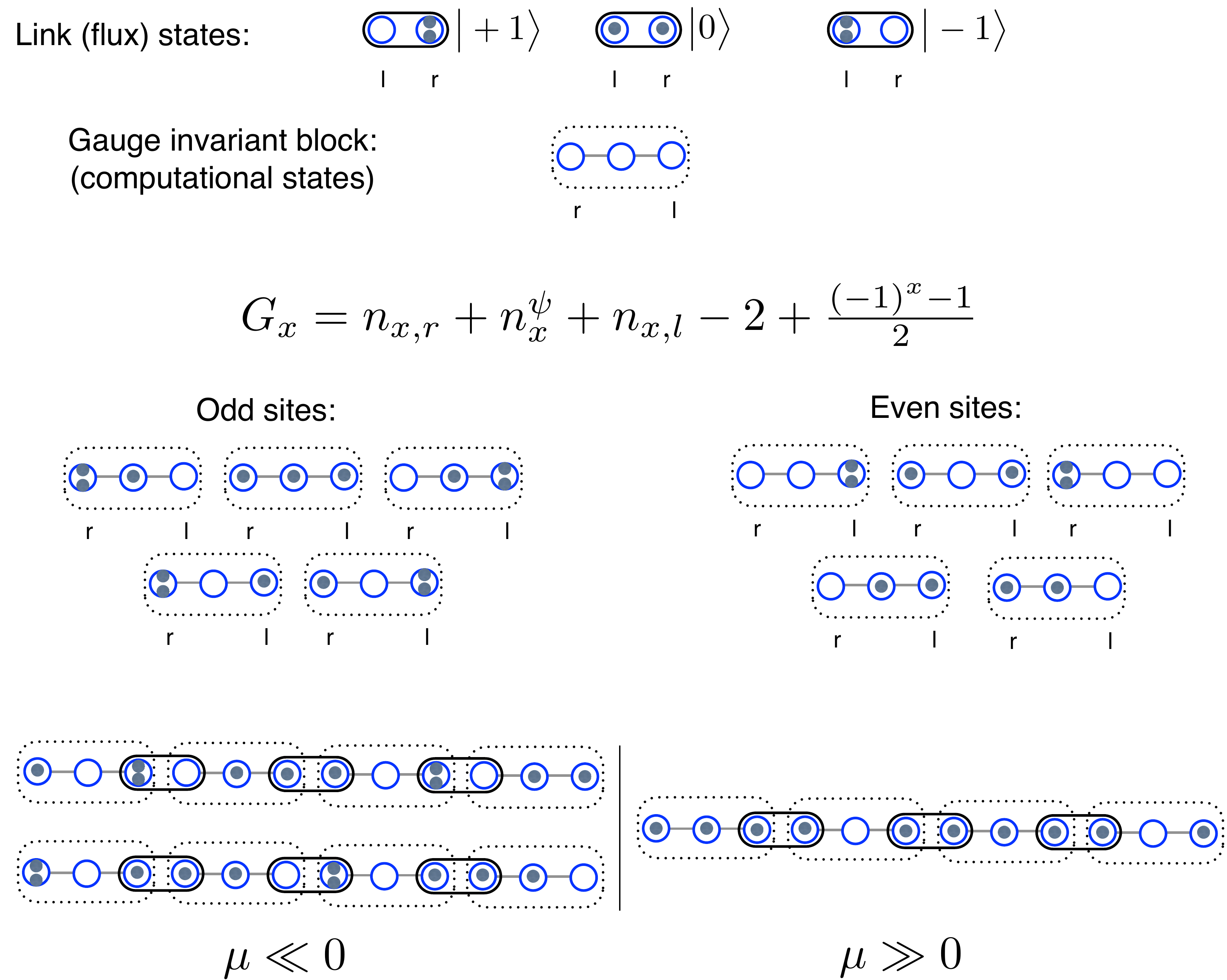}
\begin{caption}{\label{app4} Fermion and boson configurations. C and P invariant candidate vacuum state of the spin-$1$ model for $\mu \gg 0$. Competing vacua of the spin-$1$ model, which are C and P partners of each other for $\mu \ll 0 $.}
\end{caption}
\end{figure}

\begin{figure*}[ht]
\includegraphics[width=\textwidth]{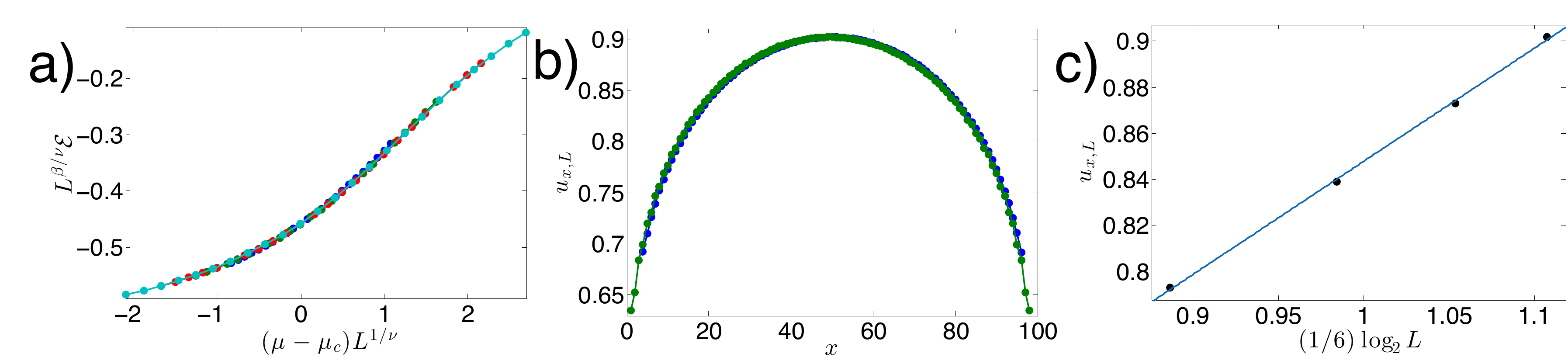}
\caption{ Results for the model with spin-$1$ on the links: a) Electric flux $\mathcal{E}$ for $L=\{40,60,80,100\}$ and $\epsilon=0.5$, with an estimate of the critical exponents $\nu \sim 1$ and $\beta \sim \frac{1}{8}$ where the overlap among the different curves is maximal. b) Uniform part of the entanglement entropy (green plot, first order approximation, i.e. $u_{x,L} = \frac{1}{2} \left( u_{x,L} + u_{x+1,L} \right)$, and blue plot third order approximation \cite{Entanglement}). c) Fit to $u_{x,L} = \frac{c}{6} \log{ \left[ \frac{2L}{\pi} \sin \left( \frac{\pi x}{L} \right) \right] }+a$, where $c = 0.49 \pm0.04$. Both, critical exponents and central charge are consistent with an Ising universality class at the phase transition. }
\label{app5}
\end{figure*}

\subsubsection{Spin-1 on the link: $N=2$} For $N=2$, the gauge invariant states are given by
\begin{equation}
\begin{split}
(2x-1): ~& | 1_{o} \rangle = |1,1,1 \rangle; ~ | 2_{o} \rangle = |2,1,0 \rangle; ~ \\
& | 3_{o} \rangle = |0,1,2 \rangle; ~ | 4_{o} \rangle = |1,0,2 \rangle; ~ \\
&| 5_{o} \rangle = |2,0,1 \rangle;  \\
(2x):  ~ &| 1_{e} \rangle = |1,0,1 \rangle; ~ | 2_{e} \rangle = |2,0,0 \rangle; ~ \\
&| 3_{e} \rangle = |0,0,2 \rangle;  ~ | 4_{o} \rangle = |1,1,0 \rangle; ~\\
& | 5_{o} \rangle = |0,1,1 \rangle;
\end{split}
\end{equation}

With these definitions, at the strong coupling limit and with $E_{0}=\frac{1}{2}$, the Hamiltonian is given by
\begin{equation}
\begin{split}
H_{\text{str}} &= \mu \sum_{x} \left( -1 \right)^{x} \psi^{\dagger}_{x} \psi_{x}  \\
&+ U \sum_{x}  \left( c^{\dagger}_{x+1,r} c_{x+1,r} + c^{\dagger}_{x,l}  c_{x,l} - 2 \right)^{2} \\
&+ \frac{1}{4} \sum_x  \left( c^{\dagger}_{x+1,r} c_{x+1,r} - c^{\dagger}_{x,l}  c_{x,l} - \left( -1 \right)^{x}  \right)^{2}
\end{split}
\end{equation}

For $\mu \ll 0 $ and due to the gauge invariance, this Hamiltonian has only two possible ground states where the configuration of the fermions is staggered 0-1-0-1 and the configuration of the bosons is also staggered by with two possible patterns: 1-2-0-1-1-2-0-1 or 2-1-1-0-2-1-1-0. This phase is two fold degenerated and the vacuum states break charge and parity symmetry.

For $\mu \gg 0$, the staggered fermion configuration is given by 1-0-1-0 and the bosons configuration is unique, given by 1-1-1-1-1-1-1-1. This phase has a unique vacuum state that is C and P symmetric.

The thermodynamical properties, phase diagram, critical behavior in the system with a spin-$1$ representation on the links at a background electric field $E_{0}=\frac{1}{2}$ is equivalent to the model with spin-$1/2$ link representation. In  Fig. \ref{app5}, we can see one particular instance where the value of the hopping term $\epsilon=0.5$, at this value the estimate of the critical point $\mu_{c}=-0.2173\pm 0.0005$ and the central charge $c = 0.49 \pm0.04$.

\end{document}